# Closed-loop recycling of rare liquid samples for gas-phase experiments

K. Fehre[*,1], M. Pitzer[*], F. Trinter[*,⸸], R. Berger[‡], A. Schießer[#] H. Schmidt-Böcking[*], R. Dörner[*] and M. S. Schöffler[*,1]

[*]Institut für Kernphysik, Goethe-Universität Frankfurt, Max-von-Laue-Str. 1, 60438 Frankfurt, Germany
[⸸]Molecular Physics, Fritz-Haber-Institut der Max-Planck-Gesellschaft, Faradayweg 4-6, 14195 Berlin, Germany
[‡]Philipps-Universität Marburg, Hans-Meerwein-Straße 4, 35039 Marburg, Germany
[#]Mass Spectrometry, Department of Chemistry, Technische Universität Darmstadt, Alarich-Weiss-Straße 4, 64287 Darmstadt, Germany

Many samples of current interest in molecular physics and physical chemistry exist in the liquid phase and are vaporized for the use in gas cells, diffuse gas targets or molecular gas jets. For some of these techniques the large sample consumption is a limiting factor. When rare, expensive molecules, such as chiral molecules or species with isotopic labels are used, wasting them in the exhaust line of the pumps is a quite expensive and inefficient approach. Therefore, we developed a closed-loop recycling system for molecules with vapor pressures below atmospheric pressure. Once filled, only a few valves have to be opened or closed and a cold trap must be moved. The recycling efficiency per turn exceeds 95 %.

**Introduction**

Experiments on single molecules in the gas phase are particularly clean and powerful as the systems are isolated from any perturbing environment. Most of the simple molecular targets, such as $H_2$, $N_2$, $O_2$, or CO, which are naturally gaseous, have extensively been investigated in the past 20 years [1–8]. In the recent past, experimental techniques improved dramatically [9–11] and research shifted the focus on more complex targets. The drawback of "larger", especially bio-molecules is that they are liquid at room temperature; many of them pose health risks or are problematic for vacuum pumps. Especially enantiopure compounds from the large group of chiral molecules, which became of high interest in molecular physics [12, 13] are usually quite expensive[14]. This is also the case for isotopically labelled species which are often custom-made [9].

Therefore, there is an effort to design the experimental setup in a way that reduces the sample consumption and minimizes contact with it. Utilizing modern detection techniques, such as velocity map imaging [15, 16], hemispherical analyzers [17], magnetic bottle spectrometers [18, 19], time-of-flight tubes [20], photon detection [21], or COLTRIMS-type reaction microscopes [22–28], at first the sample has to be introduced into the vacuum. While only a small fraction is really "used", the major part has to be pumped out of the apparatus in order to maintain the high or ultrahigh vacuum. Thus, the major part of the sample is thrown away if it is not recycled. Gas cells, effusive targets [29–32], and supersonic gas jets are the most widely used target environments. We developed a closed-loop gas-recycling system to be used for these gas environments. We have characterized it in connection with a free molecular gas jet, which is commonly used in COLTRIMS reaction microscopes.

We briefly discuss the key elements of a typical gas jet. The sample expands with its vapor pressure through a tiny nozzle (typically 30 – 200 µm) forming a supersonic gas jet when entering the vacuum recipient (referred to as expansion chamber). The expansion chamber is connected to a vacuum chamber in which the interaction takes place or to a series of differential pumping stages by a skimmer to keep the pressure in the interaction chamber as low as possible. In the interaction chamber, only a minuscule portion of the sample is investigated in the interaction with single photons, strong laser fields or charged particles and the great majority of the molecules are guided out of the reaction chamber and are differently pumped. From the typical pressures and typical pumping speeds follows that >99.9 % of the total gas load occurs in the first stage of the expansion chamber, where the gas nozzle is located. Capturing and recycling this majority of unused sample will increase efficiency dramatically.

An easy approach to recycle the sample is a simple cold trap in the fore-vacuum line between the exhaust of a high-vacuum turbomolecular pump and a dry, oil-free fore pump, for instance a roots-scroll pump combination. The "used" gas is deposited on the cold surface of the trap and transferred back into the initial reservoir for further use. This approach has some disadvantages: The gas line has to be physically opened and closed from time to time with the typical risks of leaks. The cold trap needs to defrost at a safe location (e.g., fume hood) where the sample can be transferred back into the initial reservoir or safely disposed. All this consumes valuable time and even more importantly involves the danger of getting in contact with the sample. To avoid all this, we have developed a closed-loop recycling system, which is suited for samples that are liquid at room temperature; samples with vapor pressures between 10 and 600 mbar have been used so far.

---
[1] Author to whom correspondence should be addressed: fehre@atom.uni-frankfurt.de

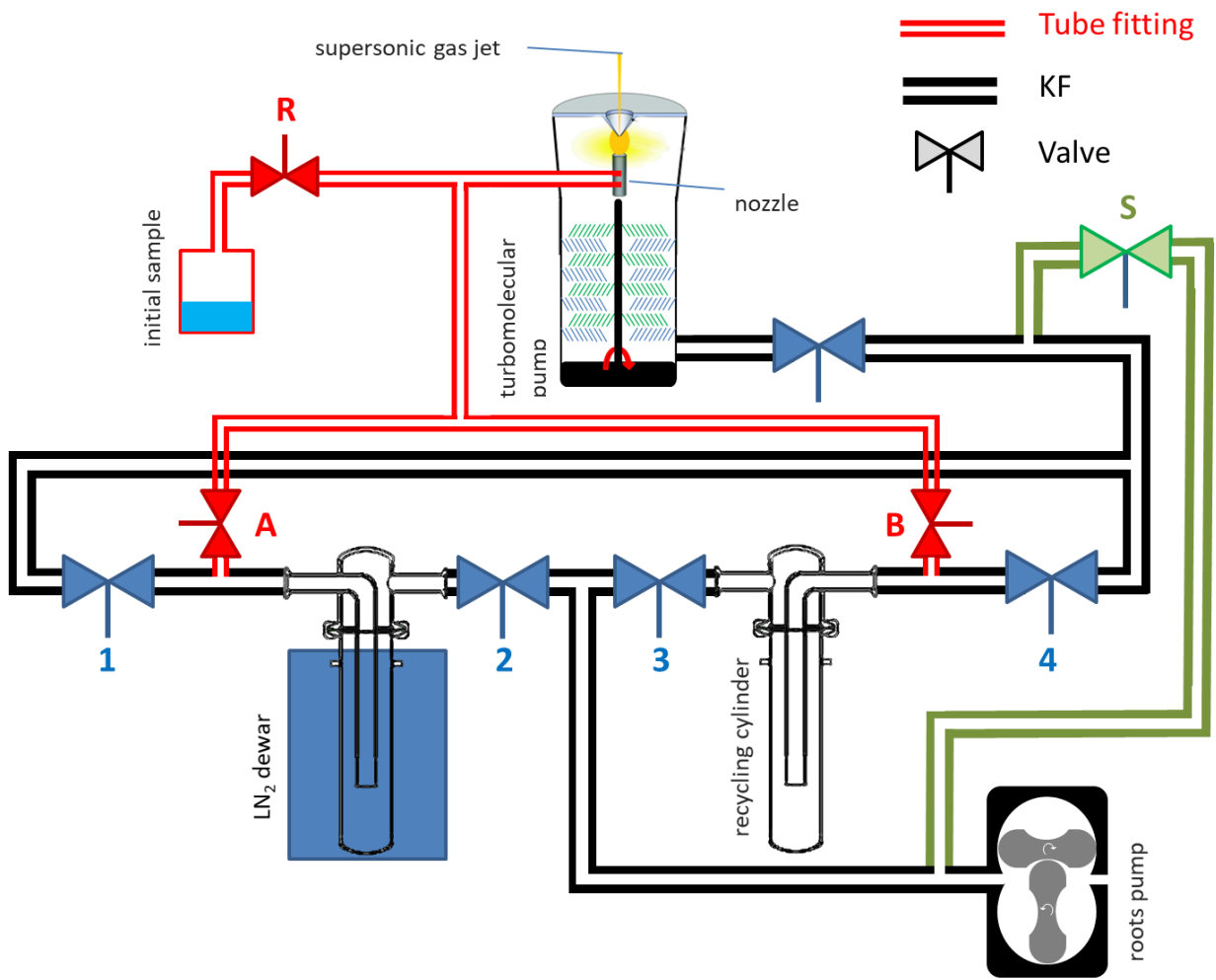

**Figure 1:** Sketch of the closed-loop sample-recycling system. A photograph is shown in Fig. 3. The sample is fed to the nozzle in tube fittings (red) from the initial sample reservoir or recycling cylinder. In the tube fittings, the pressures typically range from 10 – 500 mbar. The sample is returned from the exhaust of the turbomolecular pump to the recycling cylinder via KF pipes (black). In the KF fore-vacuum, the pressures range from $10^{-2} - 10^{-4}$ mbar. A connection made of KF pipes makes it possible to bypass the gas recycling and connects the exhaust of the turbomolecular pump directly with the pre-vacuum pump (green).

Figure 1 shows a sketch of the gas/vacuum lines, which are the main part of the recycling. A photograph is shown in Fig. 3. The supply system (e.g., in 6 mm stainless steel tubing with standard tube fittings), sketched in red connects the reservoir or one of the two cold traps with the nozzle. In black and green, the fore-vacuum system (ISO-KF25 and ISO-KF40) is sketched from the outlet of the turbomolecular pump (in the expansion chamber) via the cold traps to the fore-pump. Initially the sample is taken from the original sample cylinder (by means of valve R); valves 3 and 4, the safety bypass S and the nozzle feeding lines A and B are closed (see Fig. 2(A)). While the sample expands through the nozzle, >99.9 % are pumped with a turbomolecular pump. In the fore-vacuum line, the gas passes through a cold trap made from glass. The trap is cooled in a liquid nitrogen filled dewar flask. The sample freezes out in the cold trap. This process is so efficient that when running the pure sample, the fore pump could be switched off. Obviously when carrier gas is used (such as helium), the fore pump still has to run.

When the original sample reservoir is empty, the first recycling cycle is started. The sample is now located in the left of the two cold traps that will now be used as the new reservoir from which the sample evaporates. The dewar flask has to be removed and put underneath the other cold trap, in which for this cycle the sample freezes out. The valves 3 and 4 are opened, while the valves 1 and 2 have to be closed. Additionally, valve A must be opened and valve R closed (see Fig. 2(B)) allowing the sample to reach the nozzle. The left glass cylinder acts now as reservoir and the right one as cold trap. One can speed up the warming procedure by putting a container, filled with (warm) water underneath the first cold trap. The procedure is repeated for recycling, switching all valve positions from one to the other for each turn. See Fig. 2(C) for the path in the second turn of recycling.

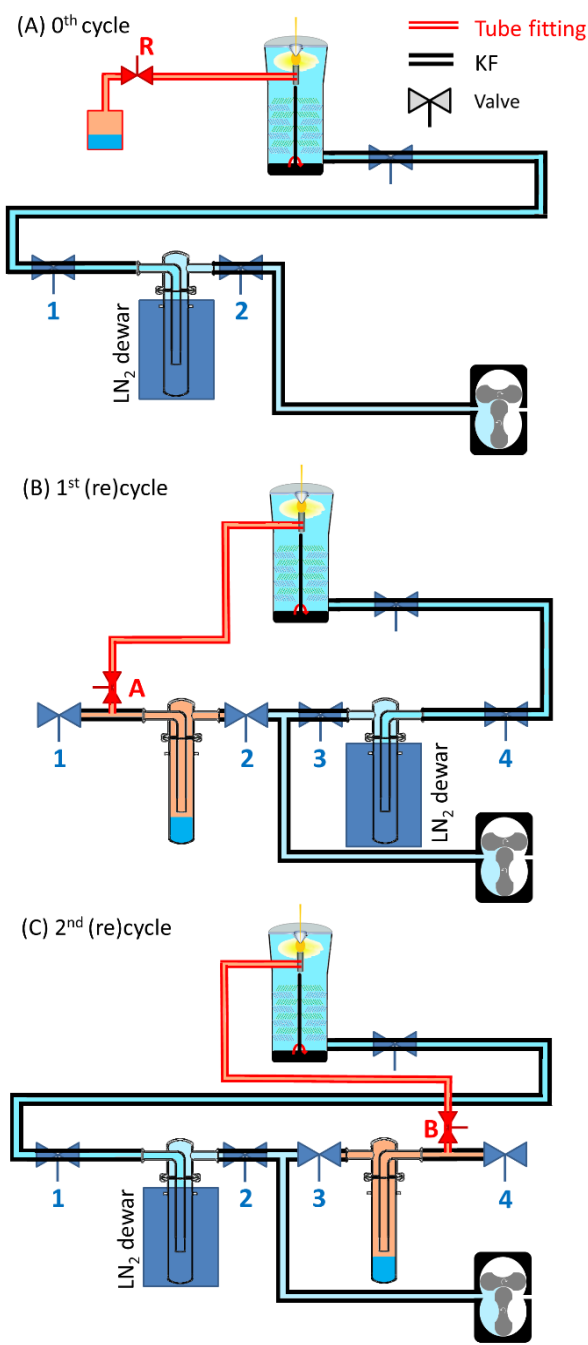

**Figure 2:** Sketch of the recycling with (A) the first filling and (B) the first and (C) the second turn on recycling mode. The former cold trap on the left-hand side acts as reservoir in the second turn, while the cylinder on the right-hand side captures the sample. Regions in the recycling system with pressures >1 (<1) mbar are highlighted in red (blue).

Care should be taken when opening the valves if, at room temperature, gaseous molecules (mostly $N_2$, $O_2$, and $CO_2$) from leaks have condensed in the cold trap. When defrosting, there is a risk that they expand quickly and the glass cylinder will burst. This never happened to us, but as a precaution, we opened the valve to the backing pump slightly at the beginning during the thawing process.

Another possible solution to this potential problem is the installation of an overpressure release valve, which opens above atmospheric pressure directly to the (toxic gas) exhaust line.

In a first version as depicted in the photograph (Fig. 3), the connectors at the glass recycling cylinders had a size of ISO-KF25 and the inner glass pipe a diameter of 15 mm. When larger quantities of liquids (>20 ml) were used, the inner glass pipe clogged by the frozen sample. This could be prevented by shortly defrosting and melting the clogged part or with a larger diameter of the inner tube of the cold trap. The latter approach was realized in a second version of this recycling system with 25 mm inner diameter of the glass pipe. As clogging still might occur, we installed a safety release valve to bypass the recycling (green line in Fig. 1). This can be opened in such a case and prevents excessive fore-vacuum pressure on the turbomolecular pump or even flooding of the expansion chamber. As protection against the potentially bursting glass of the cold traps, an acrylic glass plate on both sides protects the equipment and the user. While a cold trap made of stainless steel does not carry this risk, the glass version has the important advantage that the filling level and any contamination can be checked visually. Also it should be taken care that ISO-KF chain clamps made from plastic and ISO-KF seals with outer center ring (ideally also plastic) have to be used to avoid damage to the glass.

If samples with rather low vapor pressures are used, the complete recycling system (the fore line, the cold trap, which acts as reservoir and the supply line) have to be slightly heated. Furthermore, minimizing the surface is recommended, especially using ISO-KF pipes or tubes instead of flexible hoses, wherever this is possible; other than in our first version depicted in Fig. 3.

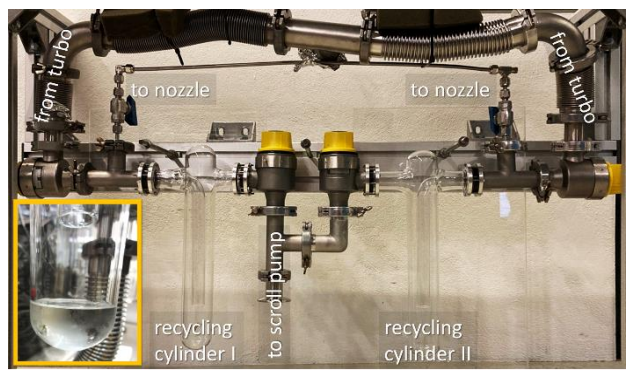

**Figure 3:** Picture for the first version of the closed-loop recycling. Cold traps (glass cylinders), ISO-KF, tubes and fittings, as well as valves and one dewar flask sum to a total of about 2 500 Euro. As can be seen in the inset, a cold trap made of glass allows the sample quantity and quality to be checked visually.

To examine the recycling efficiency, we recorded the pressure in the expansion chamber during an experiment with the closed-loop recycling system. The sample (here ~10 ml liquid trifluoro-methyloxirane with a vapor pressure of ~600 mbar, reduced with a needle valve to 100 mbar) expands through the nozzle into the expansion chamber and is pumped out by a 2300 l/s turbomolecular pump. The pressure in the expansion chamber is in very good approximation proportional to the sample flow through the nozzle. Consequently, the integral under the curves of the recycling steps in Fig. 4 gives a good estimate of how much sample was used in the respective recycling step. The properties of the supersonic jet (monitored via the event rate and the jet velocity) did not significantly change between $7.5 \cdot 10^{-5}$ and $2.7 \cdot 10^{-4}$ mbar. The integrated area under the pressure-time curve is shown in Fig. 4(B). Assuming an exponential loss of the sample, more than 95 % of the sample could be reused in each of the following recycling steps. This means instead of wasting 50 ml sample in 34 hours, only 1.5 ml out of the initial 10 ml are lost.

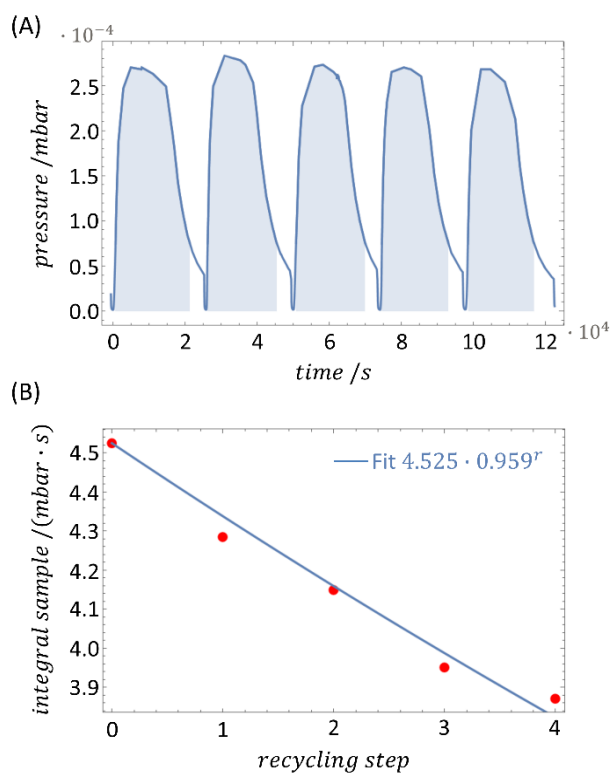

**Figure 4:** Recycling efficiency of the closed-loop recycling. (A) Pressure in the expansion chamber measured over several recycling steps. The blue areas under the curve fulfill the criterion $pressure > 7.5 \cdot 10^{-5}$ mbar separating the different recycling steps. (B) The integral of the blue areas indicate the amount of sample of every recycling step. Assuming an exponential loss of sample, more than 95 % of the sample can be reused in every recycling step.

After some recycling steps, we observed residuals (such as oil from the turbomolecular pump) on the bottom of the glass cylinders. The amount of this residual contamination could be greatly reduced by thoroughly cleaning the (re-)supply and especially the fore-line vacuum tubes. The type of sample itself also influences the residual contamination in the glass cylinder: If solvents for oil (as ethanol or methanol) were used more of the contamination was found on the bottom of the cold trap. Since only the distillate is returned to the experiment in each recycling step, the contamination did not affect our measurement. However, if the vapor pressure of the sample is very low, contaminations might interfere with the experiment. Also, in the case of very reactive samples, it is conceivable that the sample interacts with the contamination in the glass cylinder. Additionally, the mass spectra of the first and last step in the recycling were compared. This comparison did not show any noticeable change in the mass spectrum, which could be caused by any contamination of the sample by the gas recycling.

## Conclusion

In this article, we have presented the design and operation of a closed-loop recycling system for the safe handling of samples that are in the liquid phase at room temperature. With this setup, even small sample quantities can be examined in longer measurement series in the gas phase without much loss of time for the measurement or loss of sample due to difficult handling procedures. We achieved a recycling efficiency of >95 % per cycle.

Experiments with this recycling scheme have successfully been performed using methyloxirane ($C_3H_6O$) [33–35], bromo-chloro-fluoro-methane [9, 10, 36, 37], halothane [38], trifluoro-methyloxirane ($C_3H_3F_3O$), and deuterated formic acid (CHOOD) [39], with vapor pressures from 10 to 600 mbar.

## Acknowledgement

We acknowledge support from Deutsche Forschungsgemeinschaft via Sonderforschungsbereich 1319 (ELCH). M.S.S. thanks the Adolf-Messer foundation for financial support. K.F. acknowledges support by the German National Merit Foundation. We acknowledge support by Frank Scholz (DESY) in archiving the pressure as a function of time.

## Data Availability
The data that support the findings of this study are available from the corresponding author upon reasonable request.